\documentclass[onecolumn,sort&compress,numbers]{els-mrw} 

\usepackage{amsmath,amssymb,amsfonts,amsthm,makeidx,graphicx}
\usepackage{txfonts}
\usepackage{helvet}
\usepackage{siunitx}
\usepackage{subfigure}
\usepackage{cleveref}

\begin{document}


\chapter{Particle Astrophysics with High and Ultrahigh Energy Neutrinos}

\author[1]{Ke Fang}%
\author[2,3]{Kohta Murase}%

\address[1]{\orgname{University of Wisconsin--Madison}, \orgdiv{Wisconsin IceCube Particle Astrophysics Center}, \orgaddress{222 W. Washington Ave., Suite 500, Madison WI 53703}}
\address[2]{\orgname{The Pennsylvania State University}, \orgdiv{Department of Physics; Department of Astronomy and Astrophysics; Institute for Gravitation and the Cosmos}, \orgaddress{University Park, Pennsylvania 16802, USA}}
\address[3]{\orgname{Kyoto University}, \orgdiv{Yukawa Institute for Theoretical Physics}, \orgaddress{Kyoto, Kyoto 606-8502, Japan}}

\articletag{Chapter Article tagline: update of previous edition, reprint.}

\maketitle

\begin{abstract}[Abstract]
We summarize recent results of the observations of high (1~TeV-100~PeV) and ultrahigh ($\geq 100$~PeV) energy neutrinos, including the detection of a diffuse cosmic high-energy neutrino background, the identification of the first neutrino source candidates, and the observation of high-energy neutrinos from the Galactic plane. These findings open a new window to the universe by enabling the use of neutrinos to probe the cosmos that are otherwise inaccessible via photons. Although the origins of most detected neutrinos remain uncertain, we highlight several distinctive features of their sources that have emerged from current observations.  
\end{abstract}

\begin{keywords}
 	 \sep particle astrophysics \sep multi-messenger astronomy \sep neutrino astronomy 
\end{keywords}


\section*{Objectives}
\begin{itemize}
	\item Summarize recent developments in high and ultrahigh energy neutrino observations 
	\item Highlight the advances in particle astrophysics and multi-messenger astronomy enabled with these neutrinos 
\end{itemize}



\section{Introduction}  

One of the most persistent and profound mysteries in particle astrophysics is the origin of cosmic rays, especially at the highest energies. While cosmic rays have been observed over more than ten orders of magnitude in energy, the properties of the astrophysical accelerators responsible for producing the most energetic particles in the universe remain largely unknown. The fundamental difficulty in identifying their sources arises from the charged nature of cosmic rays: their trajectories are intervened by magnetic fields on Galactic and extragalactic scales. Tracking the cosmic accelerators using neutral particles, namely, photons and neutrinos produced by the cosmic rays, is therefore crucial.

High-energy neutrinos provide a complementary and potentially decisive probe of the high-energy universe. They are produced via hadronic processes, such as proton-proton or proton-photon interactions, when cosmic rays interact with surrounding matter or radiation. Because neutrinos are neutral and only weakly interact with matter and radiation, once produced, they propagate through space almost entirely unimpeded by magnetic fields or intervening matter, retaining both directional and spectral information. Their detection thus provides a powerful means to probe the environments where cosmic-ray acceleration and hadronic interactions occur.

The completion of the IceCube Neutrino Observatory at the South Pole in 2010 marked a major milestone in experimental astroparticle physics. For the first time, a detector sensitivity reached the level of astrophysical neutrino fluxes predicted by many theoretical models. In 2012–2013, IceCube reported the first compelling evidence for a diffuse flux of high-energy extraterrestrial neutrinos, marking the birth of high-energy neutrino astrophysics \citep{Aartsen2013i,Aartsen:2013jdh}. Since then, the search for the sources of these neutrinos has become a central focus of the field. 

A key development came in 2017, when IceCube detected a high-energy neutrino event (IceCube-170922A) that was found to be spatially and temporally coincident with a flaring blazar, TXS 0506+056~\citep{IceCube:2018cha}. A broad campaign of multiwavelength observations involving Fermi-LAT, MAGIC, and numerous other telescopes provided the first hint linking a specific astrophysical object to high-energy neutrino emission. This result paved a way to coordinated, neutrino-triggered follow-up observations to identify cosmic-ray accelerators. 

Since then, stronger evidence has emerged for other types of sources. In 2022, the IceCube Collaboration reported a significant excess of neutrino emission from the direction of the nearby active galaxy NGC 1068, a well-studied Type II Seyfert galaxy~\citep{IceCube:2022der}. Unlike TXS 0506+056, NGC 1068 does not have powerful jets. Its neutrino flux is orders of magnitude higher than its gamma-ray flux, suggesting that the gamma rays co-produced with the neutrinos are attenuated within the source. This makes NGC~1068 a hidden neutrino emitter and supports the idea that the all-sky neutrino flux is produced by a population of such gamma-ray-obscured neutrino sources.

More recently, in 2023, IceCube reported evidence for a neutrino excess correlated with the Galactic plane~\citep{IceCube:2023ame}. The finding suggests that Galactic sources such as supernova remnants, pulsar wind nebulae, or cosmic-ray interactions with the interstellar medium contribute about $10\%$ of the all-sky neutrino flux. 

The study of high-energy (1~TeV-100~PeV) neutrinos has been extended to even higher energies ($\geq 100$~PeV) through the development of new detection techniques and larger detectors. Although no firm detections have been made at these  energies, a few intriguing individual events have been observed.

\begin{figure*}[t!]
\centering
\includegraphics[width=0.9 \textwidth]{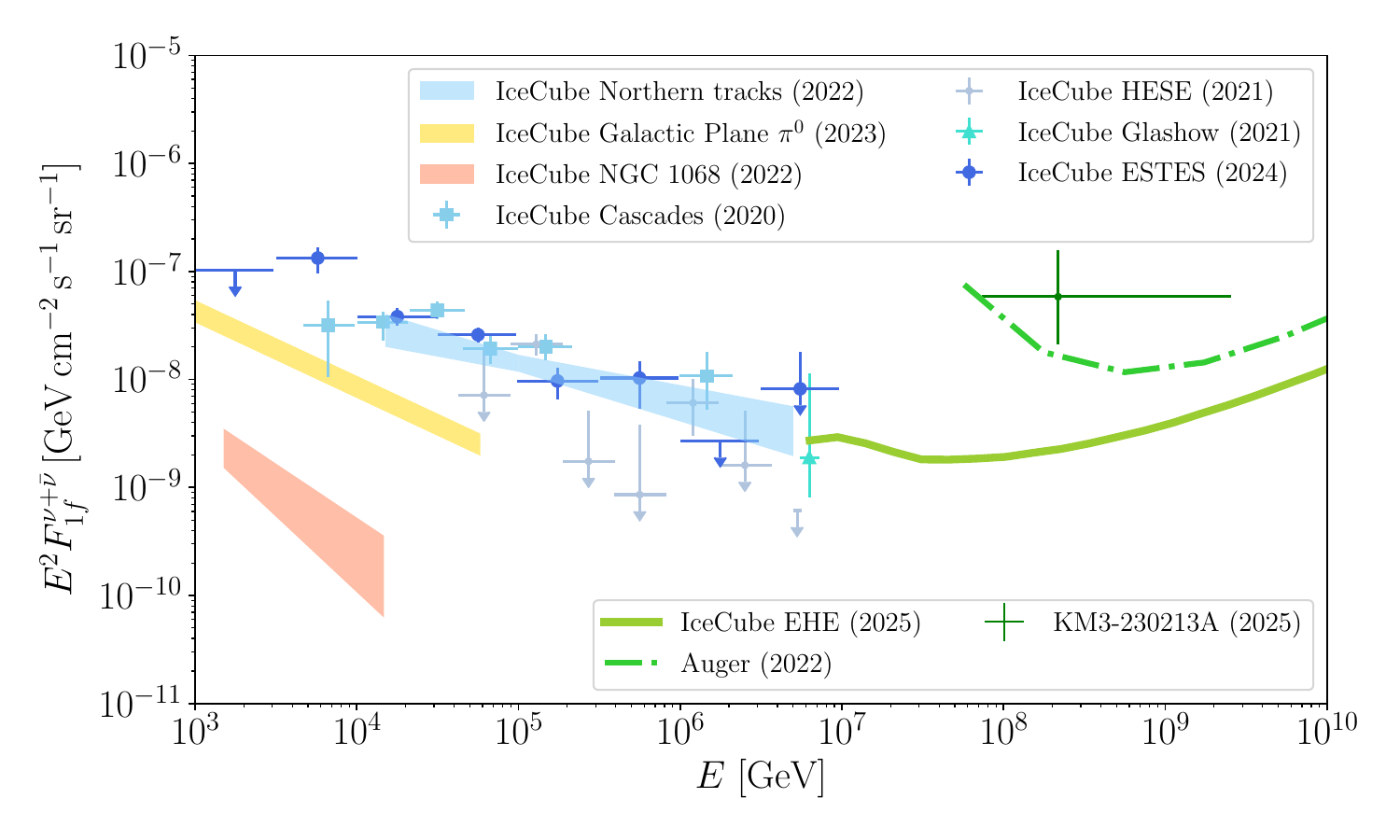}
\caption{ 
\textbf{Current observational landscape of high and ultrahigh energy  neutrinos.}  
Blue data points and shaded regions represent measurements of the diffuse astrophysical neutrino flux from various IceCube datasets~\citep{IceCube:2020wum, Abbasi:2021qfz, IceCube:2021rpz, IceCube:2024fxo}. Also shown are detections of individual sources by IceCube (scaled by $4\pi$), including the Galactic plane~\citep{IceCube:2023ame} (yellow band) and the Seyfert galaxy NGC~1068~\citep{IceCube:2022der} (red band). Green curves indicate upper limits on the diffuse UHE neutrino flux from IceCube~\citep{IceCube:2025ezc} and  Auger~\citep{AbdulHalim:2023SN}. A green marker denotes the UHE neutrino candidate event reported by KM3NeT~\citep{KM3NeTEvent}.
} 
\label{fig:sed} 
\end{figure*}

Figure~\ref{fig:sed} summarizes the current observational landscape of high- and ultrahigh-energy (UHE) neutrinos. Blue points and shaded regions show the diffuse neutrino fluxes measured by IceCube. Yellow and red bands highlight specific sources identified by IceCube. Green curves and markers represent the current limits on UHE neutrino fluxes from IceCube and the Pierre Auger Observatory, along with a recent candidate event reported by KM3NeT.

The rise of high-energy neutrino astrophysics has established neutrinos as a key component of Multimessenger Astronomy. Neutrinos now complement photons (from radio to gamma rays), cosmic rays, and gravitational waves in probing the physical conditions of astrophysical sources from active galactic nuclei (AGNs), gamma-ray bursts (GRBs), tidal disruption events (TDEs), and gravitational wave sources such as compact binary mergers. The interplay between neutrino and electromagnetic signals is essential for understanding the source physics of cosmic-ray accelerators and for probing fundamental questions about neutrino properties, and physics beyond the Standard Model.

This review aims to provide a comprehensive overview of the current state and future directions of neutrino astrophysics. We begin by discussing diffuse neutrino observations and their implications in Section~\ref{sec:diffuse}. Section~\ref{sec:extraGal} covers extragalactic neutrino sources, with a focus on recent discoveries involving TXS 0506+056 and NGC 1068. Section~\ref{sec:gal} highlights recent results on neutrino emission by the Galactic plane. Finally, we summarize the current progress and outlook for ultrahigh-energy neutrino detection in Section~\ref{sec:UHE}.

\section{All-Sky Neutrino Flux} \label{sec:diffuse}

\begin{figure*}[t!]
\centering
\includegraphics[width=0.45 \textwidth]{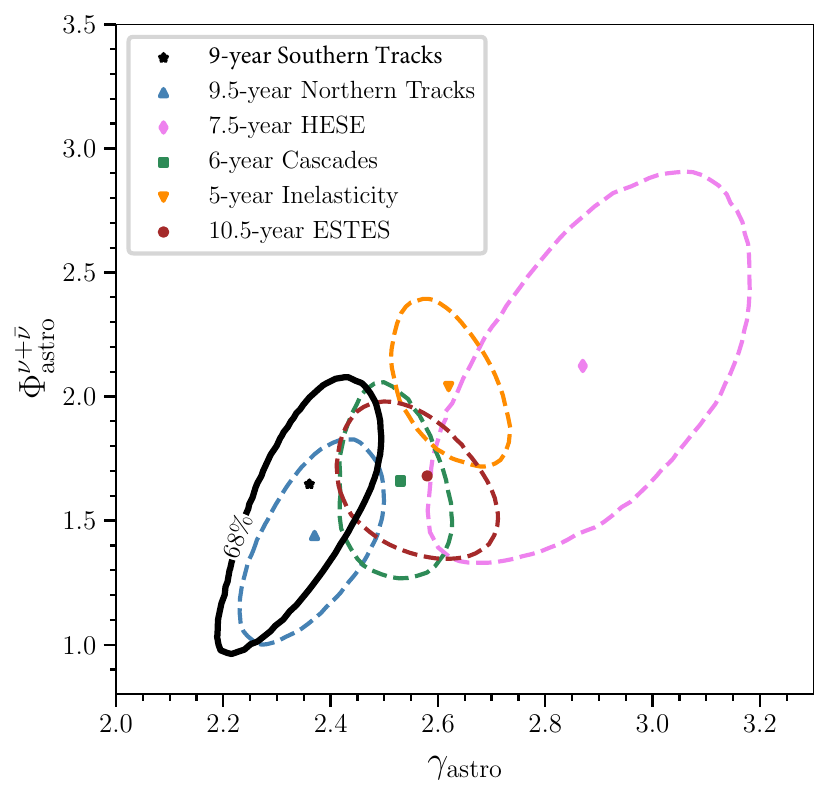}
\includegraphics[width=0.45\textwidth]{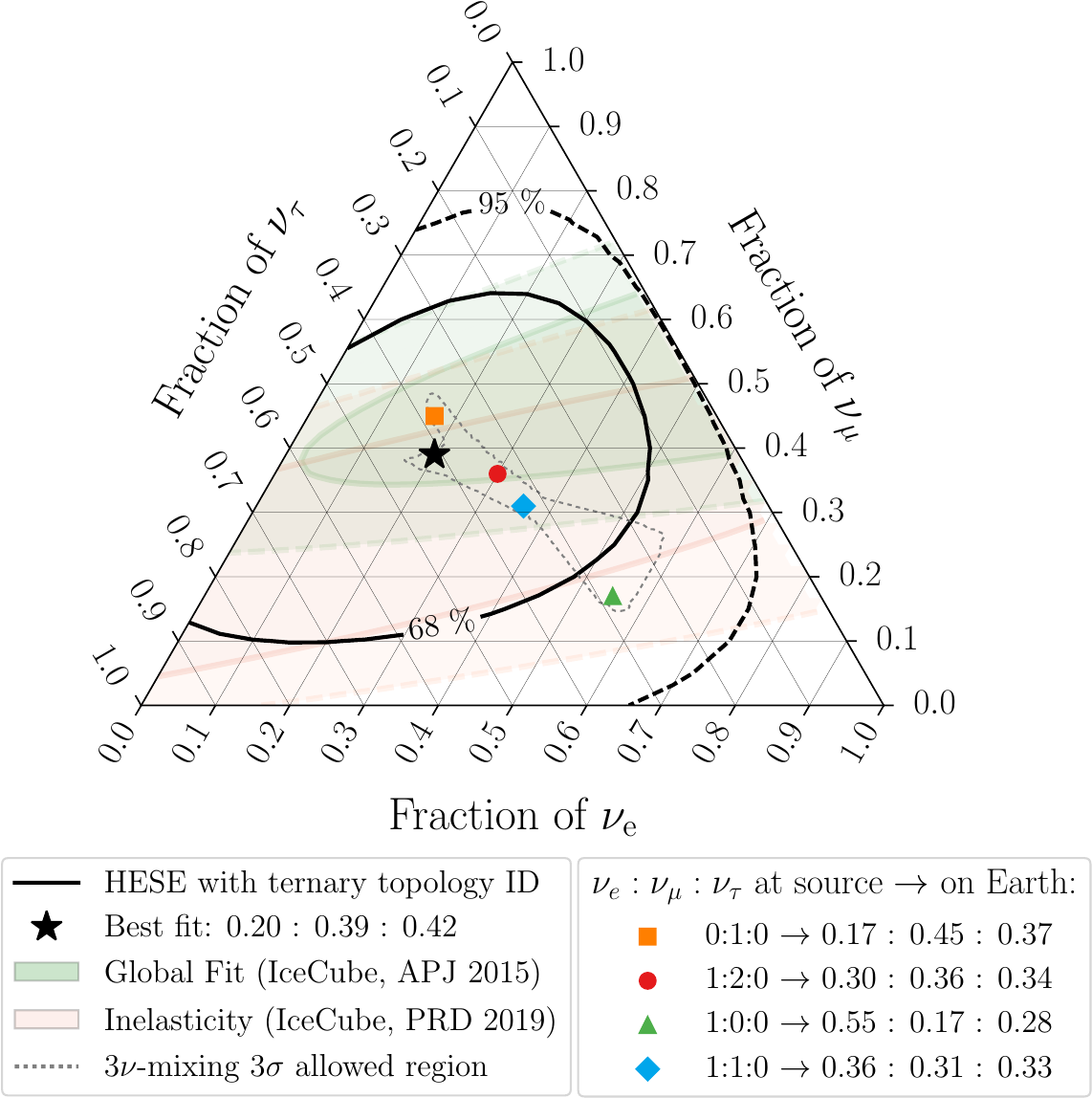}
\caption{Left: {\bf Best-fit single power-law parameters, including the per-flavor normalization and spectral index (solid markers) and 68\% contours obtained from analyses using various data samples}. Plot is from Ref.~\cite{IceCube:2025ary}. Right: Measured flavor composition of IceCube data comparing to various neutrino production models after propagation (colored markers). Contours show the $1\sigma$  and $2\sigma$ confidence intervals of the measurements. Plot is from Ref.~\cite{IceCube:2020fpi}.} 
\label{fig:diffuseParameters} 
\end{figure*}

The existence of high-energy neutrinos of astrophysical origins has been established shortly after the launch of the IceCube Observatory \citep{Aartsen2013i,Aartsen:2013jdh}. After more than ten years of IceCube operation, the flux of the diffuse astrophysical neutrinos has been measured using neutrino event samples obtained by a variety of selection criteria \citep{inelasticity5yr, HESE75yr, cascade6yr, numu95yr, IceCube:2024fxo,IceCube:2025ary}. 
The spectrum is consistent with a single power-law, 
\begin{equation}
    \Phi_\nu = \Phi_{\rm astro} \left(\frac{E_\nu}{100 \, {\rm TeV}}\right)^{-\gamma_{\rm astro}}
\end{equation}
with an index $\gamma_{\rm astro}\sim 2.4-2.9$ and a per-flavor flux of $\Phi_{\rm astro} = (1-3)\times 10^{-18}\, \rm GeV^{-1}cm^{-2}s^{-1}sr^{-1}$ at $100$~TeV as summarized in the left panel of Figure~\ref{fig:diffuseParameters}. Alternative spectral models, including a double power-law with hard and soft components, have been tested against IceCube data. While there is no strong evidence for an additional component, a combined fit to IceCube's cascade and through-going track samples shows a deviation from a single power-law, with the spectrum appearing harder at lower energies and softer at higher energies \citep{Naab:2023xcz}.

A mild excess above the atmospheric backgrounds with a similar index has also been found in the data obtained by the ANTARES \citep{Fusco:2019mN} and  Baikal Gigaton Volume Detector (Baikal-GVD) neutrino telescopes \citep{Baikal-GVD:2022fis}.

%

%
%

The diffuse neutrino flux is interestingly high when comparing to the diffuse gamma-ray flux. 
Both neutrinos and gamma rays are produced in a proton interaction. While neutral pions decay into two gamma rays, $\pi^0\to\gamma+\gamma$, the charged pions decay into three high-energy neutrinos ($\nu$) and antineutrinos ($\bar\nu$) via the decay chain $\pi^+\to\mu^++\nu_\mu$ followed by $\mu^+\to e^++\bar\nu_\mu +\nu_e$. Fluxes of gamma rays and muon neutrinos from the same interaction process  are therefore connected via a simple relation:  

\begin{equation}\label{eqn:E2dNdE_gamma_nu}
    E_\gamma^2 \frac{dN_{\gamma}}{dE_\gamma} \approx \frac{4}{K_\pi }\, \frac{1}{3} E_\nu^2 \frac{dN}{dE_\nu}  \Big| _{E_\nu = {E_\gamma}/2},
\end{equation}
where $K_\pi$ is the ratio of charged and neutral pions produced, with $K_\pi \approx 2 (1)$ for $pp (p\gamma$) interactions.

If the neutrino sources are transparent to gamma rays, the gamma rays co-produced with the neutrinos will undergo electromagnetic cascading in the extragalactic background light (EBL) and the cosmic microwave background (CMB), reprocessing their energies down to the GeV–TeV range. 
Murase et al~\citep{Murase:2013rfa} first investigated multi-messenger implications, showing the importance of 10-100~TeV neutrino measurements to probe the origin of IceCube neutrinos.  
The left panel of Figure~\ref{fig:MMconnection} presents the diffuse neutrino fluxes and those of their accompanying gamma-ray counterparts. Specifically, the pionic gamma-ray flux is computed assuming that the sources follow the star-formation (SFR) history in a flat $\Lambda$CDM universe \citep{2006ApJ...651..142H} and using equation~\ref{eqn:E2dNdE_gamma_nu} and adopting recent neutrino flux measurements over the relevant energy range.  
The figure shows that the predicted pionic gamma-ray flux exceeds the diffuse isotropic gamma-ray background, especially the non-blazar component, if the minimum neutrino energy is below $\sim 10$~TeV.  This tension suggests that gamma-ray absorption must occur at the sources, implying hidden cosmic-ray accelerators, that is, the neutrino emitters that are opaque to GeV-TeV gamma rays \cite{Murase:2015xka,2020PhRvD.101j3012C, 2021JCAP...02..037C,Fang:2022trf}. Murase et al.~\cite{Murase:2019vdl} and Fang et al.~\cite{Fang:2022trf} further suggest the cascade gamma rays from neutrino sources such as AGNs may show up at MeV to GeV energies, depending on the radiation background of the neutrino sources, and contribute to the MeV-GeV gamma-ray background. 

In addition to flux measurements, flavor composition of astrophysical neutrinos has been constrained with IceCube \citep{IceCube:2015rro,IceCube:2018pgc,IceCube:2015gsk,IceCube:2020fpi,IceCube:2024nhk}. As shown in the right panel of Figure~\ref{fig:diffuseParameters}, current constraints are compatible with astrophysical production scenarios such as pion decay and muon damp models and standard
neutrino oscillations on their propagation to Earth, while the beta-decay scenario is disfavored \citep{Bustamante:2019sdb, IceCube:2020fpi}.

\section{Extragalactic source searches}  \label{sec:extraGal}


Observationally, source searches have been carried out using various techniques, including blind searches for event clustering hot spots and  source association studies using various astrophysical source catalogs.  
Time-integrated source association searches have been performed in Northern, Southern and full sky using  IceCube's track-like events \cite{Aartsen:2019fau}, shower events \citep{IceCube:2017der}, as well as event sample from the Enhanced Starting Track Event Selection (ESTES) \citep{IceCube:2025zyb}.
Time-dependent searches have also been carried out in various forms, including the search for neutrino doublets and triplets \citep{IceCube:2025qjb} and searches for flaring gamma-ray sources \citep{IceCube:2020nig}. 
After accumulating a decade of data with a detector with gradually improved sensitivity, the first high-energy neutrino sources emerged in the neutrino sky. In light of these observations, we review below the recent progress in theoretical studies of neutrino sources across various source classes.

\subsection{The first neutrino source candidates} 

\subsubsection{Jet-quiet AGN} 
Most AGNs do not exhibit powerful jets. Jet-quiet AGNs, such as Seyfert galaxies and quasars  which dominate the extragalactic X-ray sky, have emerged as potential sources of high-energy neutrinos.

The idea that Seyfert galaxies could produce PeV neutrinos via photomeson interactions dates back to the late 1970s~\citep{Ber77,Eichler:1979yy}, with accretion shocks proposed as a mechanism for particle acceleration~\citep{Stecker:1991vm}. However, the detection of spectral cutoffs in the X-ray spectra of these sources has ruled out the original accretion shock model.  X-ray emission is now understood to originate from a high-temperature region near the black hole known as the corona. In addition, constraints from the observed MeV gamma-ray background have significantly reduced the allowed neutrino flux in such models~\citep{Stecker:2005hn}. More recent work has explored neutrino production in coronae powered by shocks as a possible explanation for the emission observed from NGC 1068~\citep{Inoue:2019yfs}.

X-ray emission from jet-quiet AGNs is widely attributed to the Comptonization by hot electrons in coronae. These coronae are believed to be powered through the magnetic dissipation. Magnetized, turbulent coronae may lead to magnetic reconnections and subsequent stochastic acceleration, and high-energy neutrinos can be produced by both $pp$ and $p\gamma$ interactions~\citep{Murase:2019vdl}. The Bethe-Heitler process plays an important role when the acceleration is slow, and the all-sky neutrino spectrum is predicted to be soft in the 10-100~TeV range with a spectral turnover which is consistent with the neutrino data.  

\begin{figure*}[t!]
\centering
\includegraphics[width=0.46\textwidth]{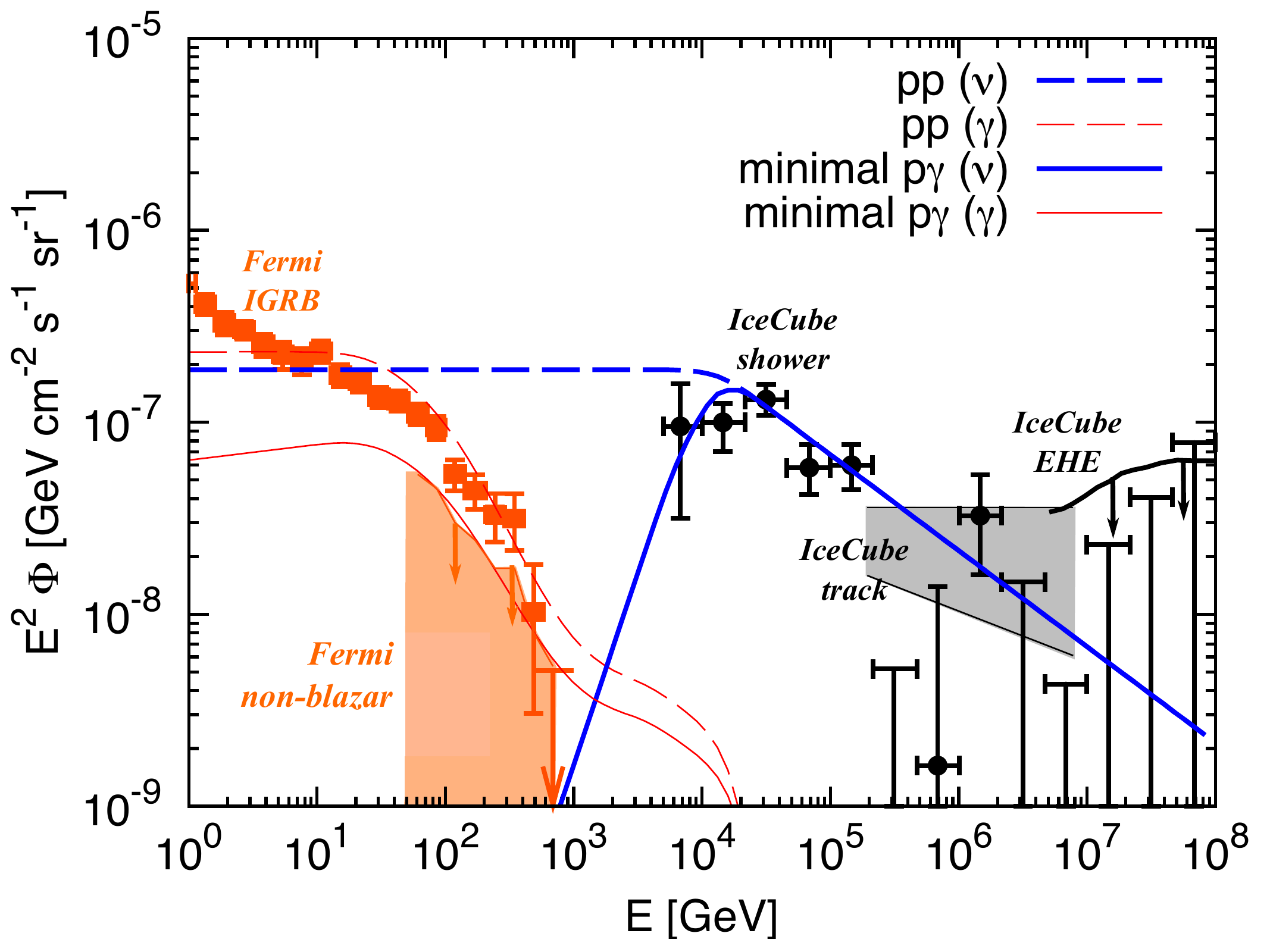}
\includegraphics[width=0.5  \textwidth]{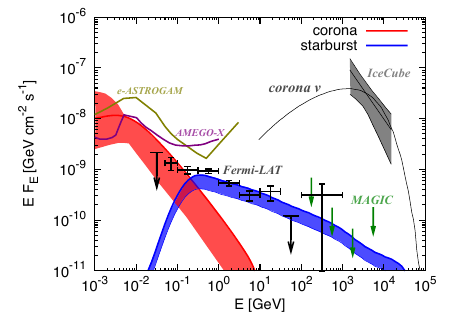}
\caption{Left: 
All-flavor neutrino (thick blue lines) and isotropic diﬀuse gamma-ray background (thin red lines) fluxes for $pp$ and $p\gamma$ scenarios, compared to IceCube and {\it Fermi}-LAT observations of the diffuse emission backgrounds~\citep{Murase:2015xka}. 
Right: Model spectra of MeV–TeV gamma-ray emission from NGC 1068 compared to {\it Fermi}-LAT data (black points)~\citep{Ajello:2023hkh}. AGN corona and starburst models are shown as red and blue shaded bands, respectively. The all-flavor coronal neutrino spectrum that explains the IceCube data (gray band) is shown by the black solid curve. Projected sensitivities of AMEGO-X and e-ASTROGAM are also shown.} 
\label{fig:MMconnection} 
\end{figure*}

Observations of high-energy neutrinos and gamma rays from NGC 1068~\citep{IceCube:2022der} also supports the prediction of hidden neutrino sources inferred by the diffuse neutrino and gamma-ray background data. The NGC 1068 neutrino flux in the 1-10~TeV range is larger than the GeV-TeV gamma-ray flux by about a factor of 10 (see the right panel of Figure~\ref{fig:MMconnection}), implying that the neutrino production should occur within $\sim15-30$ Schwarzschild radii~\citep{Murase:2022dog}. This implies the importance of understanding particle acceleration and related plasma processes in the vicinity of black holes \citep{Fang:2023vdg, Fiorillo:2023dts, Fiorillo:2024akm}.

The coronal neutrino model suggests that the neutrino luminosity is approximately proportional to the X-ray luminosity. This predicts that NGC 1068 would be the brightest source in the IceCube sky~\citep{Murase:2019vdl}. It also predicts NGC 4151 as the second-brightest source, and the Circinus galaxy as one of the brightest sources in the southern hemisphere--both consistent with observations~\citep{Murase:2023ccp}. In this model, the cascaded gamma rays eventually appear in the MeV range, which is consistent with the observation of the sub-GeV gamma-ray emission from NGC 1068~\citep{Ajello:2023hkh}.

The similar physical processes are expected for not only magentized coronae but also radiatively inefficient accretion disks, which predicts low-luminosity AGNs as promising sources of high-energy neutrinos~\cite{Kimura:2014jba,Kimura:2020thg}. They may significantly contribute to the high-energy neutrino flux in the 0.1-1~PeV range. 
Observations with next generation neutrino and MeV gamma-ray detectors will be a key to establishing jet-quiet AGNs as the sources of high-energy neutrinos. 

\subsubsection{Jet-loud AGN} 
Jet-loud AGNs are the brightest extragalactic sources in the gamma-ray sky and have long been considered among the most promising candidate sources of ultrahigh-energy cosmic rays. High-energy neutrino production in jetted AGNs, in particular their on-axis counterparts called blazars, has been widely investigated~\citep{MSB92,Mannheim:1993jg} (see also a review~\citep{Murase:2022feu}). 

A notable case is TXS 0506+056, which was identified as a neutrino source in a multimessenger campaign triggered by a neutrino of 290~TeV energy, IceCube-170922A, and supported by a retrospective identification of a neutrino burst from the same source in 2014~\citep{IceCube:2018dnn,IceCube:2018cha}. However, the interpretation remains debated as single-zone emission models have not yielded a consistent explanation. Detailed modeling of the source’s X-ray spectrum, including a distinct dip in the spectral energy distribution, has highlighted the importance of broadband data in constraining neutrino production scenarios~\citep{Keivani:2018rnh,Gao:2018mnu, MAGIC:2018sak,Murase:2018iyl, Zhang:2019dob}. While other blazars have been found in spatial coincidence with high-energy neutrino events, a physical association has not yet been firmly established.

The IceCube Collaboration has conducted stacking analyses for blazars detected by {\it Fermi}-LAT. These studies show that the blazar contribution to the all-sky neutrino flux is subdominant particularly in the 10-100~TeV range. This conclusion holds even when accounting for the contribution from unresolved blazars. Additional constraints are obtained from the lack of neutrino multiplet sources~\cite{Murase:2016gly}. These results are consistent with the finding  that the expected neutrino spectrum from blazars is too hard to explain the observed  all-sky neutrino spectrum.

Nevertheless, these results do not exclude the possibility that blazars could contribute significantly at higher energies, such as around PeV or higher. Jet-loud AGNs, including radio galaxies, also remain as compelling candidates for UHECR sources and may produce neutrinos through hadronic interactions in their surrounding environments (see Section~\ref{sec:reservoirs}).

\subsection{Other source candidates}

\subsubsection{Cosmic-ray reservoirs}\label{sec:reservoirs}
Galaxy clusters and groups are the largest gravitationally-bound objects in the universe and have been suggested as the reservoirs of cosmic rays~\citep{Berezinsky:1996wx}. Cosmic rays in these environments can be injected by AGNs and/or the intracluster gas accretion. 
Cosmic rays with $\lesssim100$~PeV can be confined in the intracluster medium for   cosmological times, and high-energy neutrinos are produced through inelastic $pp$ interactions. The predicted all-sky neutrino fluxes~\cite{Murase:2008yt,Kotera:2009ms} are consistent with the measured neutrino flux in the $\sim0.1-1$~PeV range~\cite{Murase:2013rfa}. More interestingly, 
 if AGNs are the dominant cosmic-ray sources in these environments, an astroparticle grand unification scenario becomes possible. In this framework, galaxy clusters and groups could simultaneously account for the observed sub-PeV neutrino flux, the diffuse isotropic gamma-ray background in the sub-TeV range, and the UHECR flux~\cite{Fang:2017zjf}. 
Since massive galaxy clusters are relatively rare, the scenario based on cosmic-ray acceleration at accretion shocks has been strongly constrained.

Starburst galaxies have also been considered as contributors to the all-sky neutrino flux and the diffuse gamma-ray background. The predicted flux from such galaxies~\citep{Loeb:2006tw} is broadly consistent with IceCube measurements in the 0.1–1PeV range\citep{Murase:2013rfa}. However, more recent studies indicate that their contribution to the total neutrino flux is likely modest~\citep{Sudoh:2018ana,Roth:2021lvk}. In particular, a meaningful contribution requires inclusion of starbursts hosting AGNs~\citep{Tamborra:2014xia}. If the injected cosmic-ray spectral index is approximately 2.3–2.4, the contribution from starburst galaxies to the IceCube flux is expected to be small.

\subsubsection{Gamma-ray bursts and jet-driven supernovae}
GRBs are the most powerful explosion phenomenon in the universe. High-energy neutrinos can be produced by both $pp$ and $p\gamma$ interactions. Different production sites, which include inner jet dissipation inside a star~\cite{Meszaros:2001ms,Razzaque:2004yv,Ando:2005xi,Murase:2013ffa}, inner jet dissipation outside a star~\cite{Waxman:1997ti,Bustamante:2014oka}, and external forward and reverse shocks~\cite{Waxman:1999ai,Murase:2007yt}, have been considered.

No neutrinos have been detected by IceCube and ANTARES, implying that the contribution to the all-sky neutrino flux should be less than $\sim1$\% in the TeV-PeV range~\citep{Aartsen_2017}. Nondetection of neutrinos from GRB 221009A, which is the brightest GRB ever, also gives stringent constraints on the parameter space~\citep{Murase:2022vqf}. 
The hypothesis that GRBs are the sources of UHECRs has been largely constrained although the entire parameter space has not been excluded yet. The external shock models are largely unconstrained, and future ultrahigh-energy neutrino experiments are relevant for testing the models. 

Particle acceleration in the optically-thick regions, i.e., inside the photosphere, leads to efficient neutrino production. For successful jets, photospheric neutrino emission has been considered, in which both $pp$ and $p\gamma$ components are relevant~\citep{Wang:2008zm, Murase:2008sp,Zhang:2012qy}. Jets may be choked inside the star, resulting in failed GRBs or jet-driven supernovae. However, nonthermal neutrino production is suppressed if shocks are radiation mediated, in which diffusive shock acceleration is inefficient. As a result, it is shown that slow choked jets are not promising as TeV-PeV neutrino emitters~\cite{Murase:2013ffa}. Jet-driven supernovae with extended envelope or circumstellar material can be promising hidden neutrino sources, which may account for the all-sky neutrino flux observed in IceCube~\citep{Murase:2013ffa,Senno:2015tsn}. 
Alternatively, dissipation through relativistic neutrons lead to GeV-TeV neutrinos without relying on either diffusive shock acceleration or magnetic reconnections~\cite{Bahcall:2000sa,Murase:2013hh}. Nondetection of quasithermal neutrinos from GRB 221009A disfavors neutron-proton collisions as the main dissipation mechanism~\cite{Murase:2022vqf}.  

\subsubsection{Gravitational wave sources} 
Short GRBs are widely believed to originate from compact binary mergers, including double neutron star mergers and neutron-star--black-hole mergers. The discovery of GW170817 and GRB 170817A established the former system as multimessenger sources. Short GRBs are possible high-energy neutrino emitters especially during the extended emission phase, and coincident detections of gravitational waves and neutrinos may be possible by IceCube-Gen2~\cite{Kimura:2017kan,Kimura:2018vvz}. 

Newborn magnetars that may power either supernova or merger ejecta, are among the promising sources of gravitational waves. High-energy cosmic rays may be accelerated in relativistic winds, magnetosphere or current sheets. They have been proposed as the sources of PeV-EeV neutrinos~\cite{Murase:2009pg,Fang:2013vla, Fang:2017tla} as well as quasithermal GeV-TeV neutrinos \cite{Murase:2013mpa, Carpio:2023wyt}.   

White dwarf mergers are also promising gravitational wave sources that can be detected by LISA. The magnetic dissipation and associated particle acceleration may lead to detectable high-energy neutrino emission~\cite{Xiao:2016man}.  

Supermassive black hole binary mergers are also important targets for Pulsar Timing Array and LISA~\cite{Yuan:2020oqg}. The merged black hole may launch jets, which interact with the circumnuclear material consisting of outflows from the circumbinary disk. Associated high-energy neutrino signals may be detected by next-generation neutrino detectors such as IceCube-Gen2. 

\subsubsection{Optical transients}
The gravitational core collapse of massive stars lead to supernovae, which are observed as optical transients. The majority of core-collapse supernovae are Type II SNe, and $\sim30$\% of them are Type Ibc SNe whose progenitor stars have stripped envelopes. Long GRBs are also associated with hypernovae and/or transrelativistic supernovae. Optical follow-up observations triggered by energetic neutrino events have been proposed as a method to probe such associations~\citep{Murase:2006mm,Kowalski:2007xb}.

Recent observations revealed that supernova progenitors are commonly surrounded with dense circumstellar material, which may be attributed to dense mass losses, inflated envelope, or extended chromosphere. Golden examples include SN 2013fs, 2023ixf, and 2024ggi. 
Supernovae interacting with confined circumstellar material have been proposed to be promising high-energy neutrino sources~\cite{Murase:2017pfe, Fang:2020bkm}, and detecting signals from the next Galactic supernova will bring us the multi-energy neutrino astronomy and enable us to test the hypothesis that supernovae are accelerators of cosmic rays up to the knee energy around 3~PeV. 
About $\sim10$\% of core-collapse supernovae are powered by the shock interaction with the circumstellar material, which are classified as Type IIn supernovae. Neutrinos from extragalactic Type IIn supernovae may be detectable up to $\sim5-10$~Mpc~\cite{Murase:2010cu, Partenheimer:2024qxw}. 

A star torn by a supermassive black hole is accompanied by the accretion and resulting dissipation and outflow launch. TDEs are observed as optical and X-ray transients. A small fraction of TDEs have powerful jets, which are accompanied by strong X-ray and radio emission. Neutrino emission from TDEs have been studied considering various possibilities, including jets~\cite{Wang:2011ip,Dai:2016gtz,Senno:2016bso}, disks~\cite{Hayasaki:2019kjy,Murase:2020lnu}, coronae~\cite{Murase:2020lnu}, and subrelativistic winds~\cite{Murase:2020lnu,Winter:2022fpf}.

\section{The Galactic plane}  \label{sec:gal}

%
\begin{figure*}[ht!]
\centering
\includegraphics[width=0.55\textwidth]{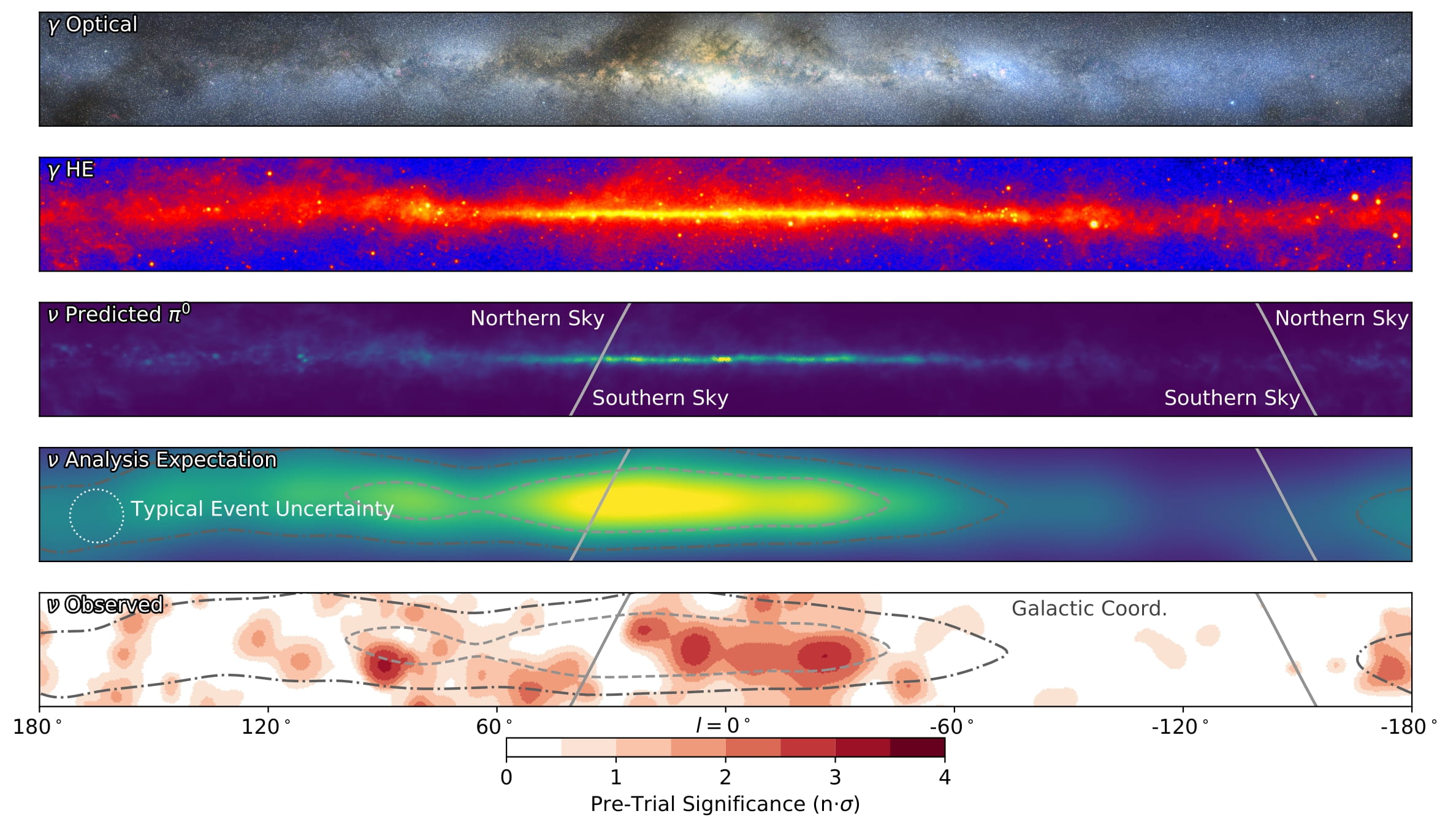}
\includegraphics[width=0.44\textwidth]{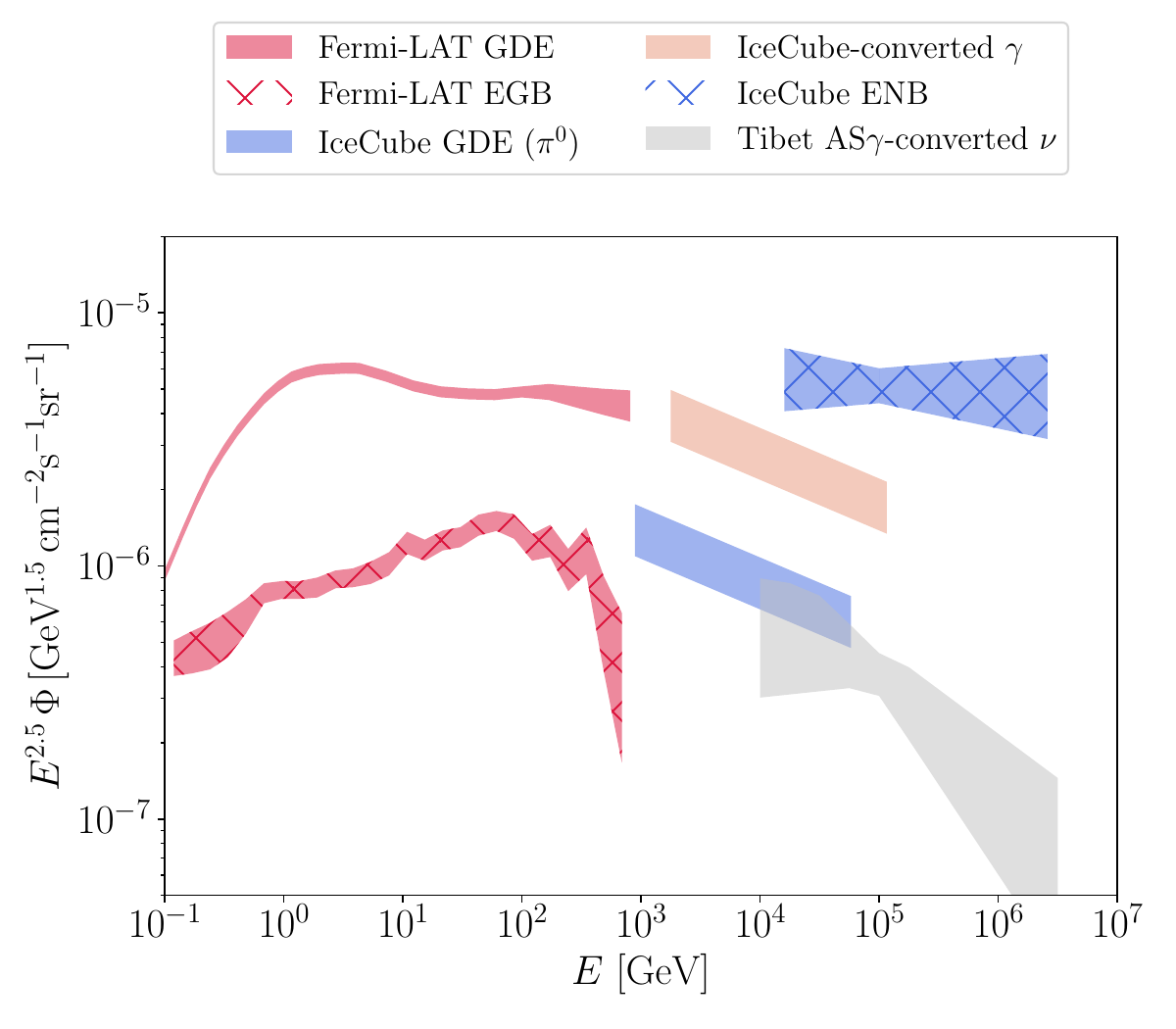}

\caption{Left (from Ref.~\cite{IceCubeGP}): {\bf The plane of the Milky Way galaxy in photons and neutrinos.} Each panel from top to bottom shows the flux of the Milky Way in latitude and longitude in the following wavelengths or messengers: (A) optical, (B)  gamma-ray based on the {\it Fermi}-LAT 12-year survey at energies greater than 1 GeV, (C) neutrino, as a template derived from the $\pi^0$ template that matches the {\it Fermi}-LAT observations of the diffuse gamma-ray emission, (D) neutrino, as an  emission template from panel (C) but including the IceCube sensitivity to cascade-like events.  The dotted white circle indicates the angular uncertainty of a typical event. Contours indicate the central regions that contain 20\% and 50\% of the predicted diffuse neutrino emission signal. (E) neutrino, as the pre-trial significance of the IceCube neutrino observations, calculated from the all-sky scan for point-like sources using the cascade neutrino event sample. Contours are the same as in panel (D). Right (from Ref.~ \cite{Fang:2023azx}): {\bf All-sky-averaged intensities, scaled by $E^{2.5}$, of the Galactic diffuse emission (GDE) and extragalactic background (EB) in gamma-ray and neutrino.} In this plot, blue and grey colors indicate neutrino measurements and red color corresponds to gamma-ray observations. Non-hatched bands indicate observations of the  Galactic diffuse emission and hatched bands correspond to observations of the extragalactic diffuse emission. Galactic components include the diffuse neutrino emission (per-flavour flux) from the Galactic plane measured by IceCube using the $\pi^0$ template \cite{IceCubeGP} and the neutrino flux derived from the gamma-ray GDE measured by
Tibet AS$\gamma$ (grey-shaded region) \cite{Tibet21} and the Galactic interstellar emission model
of Fermi-LAT (red-shaded region)\cite{4FGL}. Extragalactic components include the isotropic diffuse neutrino background measured by IceCube (blue-hatched region)\cite{Aartsen:2020aqd} and the EGB measured by Fermi-LAT (red-hatched region)\cite{FermiIGRB}.
 }
\label{fig:MultimessengerGalacticPlane} 
\end{figure*}

A population of high-energy neutrinos from the Galactic plane was identified in the IceCube data in 2023~\citep{IceCubeGP}. This discovery built on previous efforts using the IceCube data~\cite{Aab:2013ika,IceCube:2017trr,IceCube:2019lzm}, combined IceCube and ANTARES data~\cite{IceCube:2017trr,IceCube:2019lzm,ANTARES:2018nyb,ANTARES:2017nlh} and benefited from recent improvements in event reconstruction and angular resolution enabled by deep learning techniques~\cite{Abbasi:2021ryj,IceCube:2021umt,mirco_huennefeld_2022_7412035}.

The search for Galactic neutrinos uses templates of the Galactic diffuse emission. These include 1) the {\it Fermi} $\pi^0$ model, a GALPROP model \cite{Strong:1998pw}  extrapolated to the TeV range using a single power law with index $-2.7$, 2) KRA models with cosmic-ray cutoffs at 5 and 50~PeV \cite{Gaggero:2015xza}, which include radially dependent cosmic-ray transport properties, 3) the CRINGE model, fitted to recent cosmic-ray data from AMS-02, DAMPE, IceTop, and KASCADE \cite{Schwefer:2022zly}, as well as 4) simplified models derived from a convolution of source distribution models and gas maps \cite{FangMurase21}. 

Using the first two templates, ref.~\cite{IceCube:2023ame} rejected the background-only hypothesis with a post-trial significance of 4.5$\sigma$ (see the left panel of Figure~\ref{fig:MultimessengerGalacticPlane}). The search with tracks and the other templates identified a Galactic contribution in the northern sky at the $2.7\sigma$ level with a consistent flux \cite{IceCube:2023hou}.

The ANTARES collaboration also searched for neutrino emission from the Galactic ridge region, defined as $|l|<30^\circ$ and $|b|<2^\circ$ \cite{ANTARES:2022izu} and identified a mild excess at the $2.2\,\sigma$ level. The best-fit spectrum is found to be $dN/dE_\nu\propto E_\nu^{-2.45}$, suggesting a harder cosmic-ray spectrum in the inner Galaxy.  The measured neutrino flux is consistent with the model of the interaction of diffuse cosmic rays with the ISM gas in the Galactic ridge. 

Similarly, Baikal-GVD reported an excess of cascade events from low Galactic latitudes using six years of data~\citep{Baikal-GVD:2024kfx}.

The IceCube observation of the Galactic plane revealed an important fact about the Milky Way--in its current state, our own galaxy is not a typical neutrino emitter \cite{Fang:2023azx}. This is because despite of its proximity, the brightness of the Milky Way in neutrinos is not much higher than that of the extragalactic neutrino flux, as demonstrated in the right panel of Figure~\ref{fig:MultimessengerGalacticPlane}. This stands in contrast to observations in electromagnetic wavelengths, where the Galactic plane is typically the most prominent feature in the sky. Based on a Galactic flux contribution of $\sim$10\% of the total all-sky neutrino flux at 30 TeV, \cite{Fang:2023azx} estimated that the Milky Way's total neutrino luminosity is less than 1\% of that expected from an external galaxy with similar stellar mass. This suggests that certain powerful neutrino source classes such as active galaxies are absent in our Galaxy, possibly due to the long-term inactivity of the central black hole.

This aligns with pre-IceCube predictions, which suggested that the neutrino sky would be dominated by energetic extragalactic sources such as AGNs~\citep{Stecker:1991vm,MSB92}, galaxy clusters and groups~\citep{Murase:2008yt,Kotera:2009ms}, starburst galaxies~\citep{Loeb:2006tw}, and rare transients such as transrelativistic supernovae~\citep{Murase:2006mm}, while the diffuse Galactic neutrino flux is lower~\citep{Stecker:1978ah} and the contribution from Milky-Way-like galaxies is more negligible~\citep{Stecker:1979he}.

Since gamma rays from Galactic sources can reach Earth largely unattenuated, comparing neutrino and gamma-ray emission provides a powerful test of their origin. 
Early works following IceCube's 2013 diffuse flux discovery~\citep{Aartsen:2013jdh,Fox:2013oza,Razzaque:2013uoa,Ahlers:2013xia,Lunardini:2013gva} considered a range of possible Galactic sources and constrained the Galactic contribution to the all-sky neutrino flux to $\lesssim$ 10–30\%~\citep{Ahlers:2013xia,Joshi:2013aua,Murase:2015xka}.
In 2021, the Tibet AS$\gamma$ experiment reported diffuse gamma-ray emission from the Galactic plane~\citep{Tibet21}, which was later confirmed by LHAASO~\citep{LHAASO:2023gne}. These measurements, especially at sub-PeV energies, enabled the first model-independent estimates of the diffuse Galactic neutrino flux under the hadronic interaction scenario~\citep{Fang:2021ylv}. The observed similarity between the diffuse Galactic neutrino and gamma-ray fluxes supports a hadronic origin for the gamma rays~\citep{Fang:2023ffx}.

The observed Galactic neutrino emission likely results from a combination of diffuse cosmic-ray interactions and unresolved sources \citep{Fang:2024nxn}. 
In light of the IceCube Galactic plane observation, models have suggested that hadronic sources such as star-forming regions \citep{Mitchell:2024yuy, 2025arXiv250314651A}, X-ray binaries \cite{2025arXiv250117467K} and their coronae \citep{2024ApJ...975L..35F} or traditional leptonic sources such as unresolved pulsars \citep{2024ApJ...975L...6K}. Source association studies with IceCube data \cite{IceCube:2022jpz, IceCube:2022heu, IceCube:2023ame} or joint analysis HAWC and IceCube data \citep{Alfaro:2024suy} have constrained the contribution of Galactic source populations to high-energy neutrinos. 

Using $\gamma$-ray catalogs including 4FGL, HGPS, 3HWC, and 1LHAASO, and measurements of the Galactic diffuse emission by Tibet AS$\gamma$ and LHAASO, \cite{Fang:2023ffx} derives maximum fluxes of neutrino emission from the Galactic plane  through the multi-messenger connection between neutrinos and $\gamma$ rays in hadronic interactions (see equation~\ref{eqn:E2dNdE_gamma_nu}). The IceCube Galactic neutrino flux is comparable to  the sum of the diffuse gamma-ray emission and the contribution from all resolved gamma-ray sources when excluding likely leptonic emitters such as pulsars, pulsar wind nebulae, and TeV halos. Since not all gamma-ray sources are efficient neutrino emitters, it is likely that the Galactic neutrino emission is dominated by the diffuse emission by the cosmic-ray sea and unresolved hadronic $\gamma$-ray sources \citep{2023ApJ...957L...6F,2025PhRvD.111f3035L, 2025ApJ...984...98K}.

\section{Ultrahigh Energy Neutrinos}  \label{sec:UHE}
\subsection{Theory} 

Ultra-high-energy cosmic rays (UHECRs), particles with energies exceeding \(1\,\mathrm{EeV} = 10^{18}\,\mathrm{eV}\), were first detected in the 1960s~\citep{PhysRevLett.10.146} and have been  measured in detail by the Pierre Auger Observatory and the Telescope Array over the past decade. The UHECR energy spectrum exhibits a suppression above \(\sim 40\,\mathrm{EeV}\)~\citep{PierreAuger:2008rol, TelescopeArray:2012qqu, PierreAuger:2025hnw}, consistent with the Greisen–Zatsepin–Kuzmin (GZK) cutoff~\citep{Greisen1966, Zatsepin:1966jv}, which arises from interactions between UHECRs and the cosmic microwave background (CMB). These interactions define the GZK horizon, which is the maximum distance from which UHECRs can propagate to Earth before losing a significant fraction of their energy.

Photopion production in these interactions generates charged pions, which decay into ultra-high-energy neutrinos, commonly referred to as ``cosmogenic'' neutrinos. The cosmogenic neutrino spectrum is expected to peak around \(1\,\mathrm{EeV}\), although it may shift to lower energies if the UHECR composition lacks a substantial proton component above tens of EeV \citep{Kotera:2010yn, Ahlers2012}. Since photodisintegration is less efficient than photomeson production in producing neutrinos, the cosmogenic neutrino flux strongly depends on the proton fraction in the UHECRs. It also depends on the amount of sources beyond the GZK horizon. Model predictions vary widely, with fluxes as low as \(10^{-10}\,\mathrm{GeV\,cm^{-2}\,s^{-1}\,sr^{-1}}\) under composition fits based on Auger data \citep{AlvesBatista:2018zui}, and as high as \(10^{-8}\,\mathrm{GeV\,cm^{-2}\,s^{-1}\,sr^{-1}}\) when invoking additional proton components at the highest energies \citep{vanVliet:2019nse}.

Ultra-high-energy (UHE) neutrinos are also expected to originate from the sources of UHECRs themselves, as these sources may reside in environments with dense gas and intense radiation fields. Models involving such environments, such as galaxy clusters with central accelerators \citep{Murase:2008yt,Fang:2017zjf}, fast-spinning newborn pulsars \citep{Murase:2009pg,Fang:2013vla}, active galactic nuclei \citep{Murase:2014foa,Rodrigues:2020pli}, and gamma-ray burst afterglows \citep{Murase:2007yt}, can potentially account for the observed UHECR flux. These source scenarios predict that the resulting neutrino flux may be comparable to, or even exceed, the flux of cosmogenic neutrinos, making them promising targets for current and next-generation UHE neutrino observatories.


\subsection{Observations}

{\bf Detection techniques--} The study of ultra-high-energy (UHE) neutrinos remains a largely unexplored frontier, as no neutrinos with energies above 100~PeV have been definitively detected to date. Optical Cherenkov detectors remain sensitive in this energy range when deployed over sufficiently large volumes. They are also capable of detecting neutrinos that have cascaded from EeV $\nu_\tau$ down to PeV energies. Following the success of the IceCube telescope, optical neutrino detectors being built and planned include KM3NeT \citep{KM3Net:2016zxf}, Baikal-GVD \citep{Baikal-GVD:2019kwy}, P-One \citep{P-ONE:2020ljt}, IceCube-Gen2 \citep{IceCube-Gen2:2020qha}, TRIDENT \citep{TRIDENT:2022hql}, HUNT \citep{2024JInst..19T8006P}.

In parallel, alternative and generally less expensive detection strategies have been developed, which broadly fall into two categories.

The first approach seeks to detect all neutrino flavors using radio techniques, through the Askaryan radiation generated by compact electromagnetic showers from UHE neutrino interactions in ice, the atmosphere, or even space. This technique has been verified in previous experiments including ARA \citep{ARA:2022rwq, ARA:2014fyf}, ARIANNA  \citep{ARIANNA:2019scz,2020arXiv200409841A}, and ANITA \citep{ANITA:2016vrp,ANITA:2018sgj}, and implemented in on-going  and planned projects such as RNO-G \citep{RNO-G:2020rmc}, PUEO \citep{PUEO:2020bnn}, and IceCube-Gen 2 radio array \citep{IceCube-Gen2:2020qha}.  

The second approach focuses on detecting $\nu_\tau$ specifically, by targeting the extensive air showers initiated by tau leptons produced in charged-current interactions of Earth-skimming UHE $\nu_\tau$ that interact within the Earth. These air showers can be observed using established techniques such as water-Cherenkov particle detectors (as in planned experiments AugerPrime \citep{PierreAuger:2016qzd}, GCOS \citep{GCOS:2021exh}, and TAMBO  \citep{Romero-Wolf:2020pzh}), air-shower imaging telescopes (as in Trinity \citep{Otte:2018uxj} and Ashra NTA \citep{Sasaki:2014mwa}, EUSO-SPB2 \citep{Adams:2017fjh}, POEMMA \citep{POEMMA:2020ykm}) , and radio detection of air showers (as in BEACON \citep{Wissel:2020sec} and GRAND \citep{GRAND:2018iaj}; also implemented in ANITA and PUEO). More discussion regarding the detection techniques of UHE neutrinos and future experiments may be found in reference~\citep{Ackermann:2022rqc}. 

\vspace{0.5em}

{\noindent \bf Limits on flux--} Various experiments have placed upper limits on the diffuse UHE neutrino flux~\citep{ARA:2019wcf, Anker:2019rzo, ANITA:2019wyx, PierreAuger:2019ens, AbdulHalim:2023SN}. To date, the most stringent constraint comes from the analysis presented in~\cite{IceCube:2025ezc}. Using 12.6 years of IceCube data, a search for extremely high-energy neutrinos yielded no events above 10~PeV, constraining the all-flavor diffuse neutrino flux to the level of \(E^2 \phi_{\rm all\text{-}flavor} \approx 10^{-8}\,\mathrm{GeV\,cm^{-2}\,s^{-1}\,sr^{-1}}\) at \(10^{18}\)~eV. This null result places strong constraints on the composition of UHECRs, disfavoring a proton-dominated scenario above \(\sim 30\)~EeV at the 90\% confidence level, assuming that the source evolution follows or exceeds the star formation rate.

\vspace{0.5em}

{\noindent \bf Individual events--} Despite of a firmly observed population of UHE neutrinos, a few interesting events have been detected, including the ANITA anomalous events and KM3NeT UHE event KM3-230213A.

ANITA,  a balloon-borne Antarctic UHE
particle detector, typically  distinguishes between direct and ice-reflected air showers by their arrival direction and polarity. Across its four flights, ANITA detected a handful of anomalous events with arrival angles and polarities inconsistent with expected reflections \citep{ANITA:2016vrp,ANITA:2018sgj,ANITA:2021xxh}. These anomalies prompted speculation about a possible Earth-skimming tau neutrino origin, but detailed analyses show this interpretation is strongly disfavored under Standard Model assumptions and UHE neutrino limits from other experiments, including IceCube \citep{IceCube:2016uab, Romero-Wolf:2018zxt, IceCube:2020gbx} and Auger \citep{PierreAuger:2025hvl}. The origin of these anomalous events remains unclear and will be explored with follow-up observations \citep{PUEO:2020bnn}.

KM3-230213A is an approximately horizontal and extremely
 energetic muon track detected by the KM3NeT/ARCA telescope \citep{KM3NeTEvent}. The event was observed
using a conguration of 21 detection lines, which constitute
about 10\% of the planned ARCA detector. The energy and
orientation of this track suggest that it is not of atmospheric origin. The estimated muon energy is $\sim$120~PeV, inferring a neutrino energy that produces such muons at  $\sim$220~PeV. 
If this event was produced by a neutrino that is part of an isotropic, diffuse flux, or if it was from a long-term, steady source, it would be in tension with constraints that have been placed by the IceCube Neutrino Observatory~\citep{IceCube:2018fhm, IceCube:2024fxo,IceCube:2025ezc}, which has a larger effective area and has been collecting data for much longer than KM3NeT/ARCA \citep{KM3NeT:2018wnd, KM3NeT:2024uhg,Aartsen:2019fau}. Such   scenarios would thus require the KM3NeT event to be an unlikely upward fluctuation. A transient source hypothesis, where the event arised from a brief, individual source of UHE neutrinos, could reduce this tension \citep{Li:2025tqf, Muzio:2025gbr, Fang:2025nzg, IceCube:2025ezc}. 
Such events may be accompanied by electromagnetic cascades \citep{Fang:2025nzg} or synchrotron halos \citep{Murase:2011yw, Sherman:2025gir} resulted from co-produced UHE photons, but no clear GeV–TeV counterparts have yet been identified within the  $\sim 3^\circ$ localization of the KM3NeT event \citep{KM3NeT:2025bxl,2025arXiv250316606C}.


\section{Conclusions}
\label{sec:conclusions}
The past decade marked the beginning of neutrino astrophysics, with the first detection of extraterrestrial neutrinos, the identification of the first astrophysical neutrino sources, and the discovery of neutrino emission from within our own Galaxy. Important questions remain, including the origin of the bulk of IceCube's neutrinos and the existence of a neutrino population above 100 PeV. Progress in the field will rely on future detectors with improved angular resolution and flux sensitivity. These advances will make it possible to use neutrinos, together with electromagnetic waves, cosmic rays, and gravitational waves, to explore the universe.

\begin{ack}[Acknowledgments]%
K.F. acknowledges support from the National Science Foundation (PHY-2238916, PHY-2514194) and the Sloan Research Fellowship. This work was supported by a grant from the Simons Foundation (00001470, KF). 
The work of K.M. was supported by the NSF Grants No.~AST-2108466 and No.~AST-2108467, and KAKENHI No.~20H05852.
\end{ack}

\newcommand{\astropart} {Astropart. Phys.}
\newcommand {\araa}     {Annual Review of Astronomy and Astrophysics}
\newcommand {\actaa}    {Acta Astronomica}
\newcommand {\ao}       {Applied Optics}
\newcommand {\aj}       {Astronomical Journal}
\newcommand {\azh}      {Astronomicheskii Zhurnal}
\newcommand {\aap}      {A\&A}
\newcommand {\aapr}     {A\&A Reviews}
\newcommand {\apj}      {ApJ.}
\newcommand {\apjl}     {ApJ Letters}
\newcommand {\apjs}     {The Astrophysical Journal Supplement Series}
\newcommand {\aplett}   {Astrophysics Letters and Communications}
\newcommand {\apspr}    {Astrophysics Space Physics Research}
\newcommand {\apss}     {Astrophysics and Space Science}
\newcommand {\aaps}     {Astrophysics and Space Science Supplement Series}
\newcommand {\brjp}     {Brazilian Journal of Physics}
\newcommand {\baas}     {Bulletin of the AAS}
\newcommand {\bain}     {Bulletin Astronomical Inst. Netherlands}
\newcommand {\caa}      {Chinese Astronomy and Astrophysics}
\newcommand {\cjaa}     {Chinese Journal of Astronomy and Astrophysics}
\newcommand {\coasp}    {Comments on Astrophysics and Space Physics}
\newcommand {\fcp}      {Fundamental Cosmic Physics}
\newcommand {\grg}      {General Relativity and Gravitation}
\newcommand {\gca}      {Geochimica Cosmochimica Acta}
\newcommand {\grl}      {Geophysics Research Letters}
\newcommand {\iaucirc}  {IAU Circular}
\newcommand {\icarus}   {Icarus}
\newcommand {\ijmpa}    {International Journal of Modern Physics A}
\newcommand {\ijmpb}    {International Journal of Modern Physics B}
\newcommand {\ijmpc}    {International Journal of Modern Physics C}
\newcommand {\ijmpd}    {International Journal of Modern Physics D}
\newcommand {\ijmpe}    {International Journal of Modern Physics E}
\newcommand {\jcp}      {Journal of Chemical Physics}
\newcommand {\jgr}      {Journal of Geophysical Research}
\newcommand {\jcap}     {JCAP}
\newcommand {\jqsrt}    {Journal of Quantitative Specstroscopy and Radiative Transfer}
\newcommand {\jrasc}    {Journal of the RAS of Canada}
\newcommand {\memras}   {Memoirs of the RAS}
\newcommand {\memsai}   {Mem. Societa Astronomica Italiana}
\newcommand {\mnras}    {MNRAS}
\newcommand {\na}       {New Astronomy}
\newcommand {\nar}      {New Astronomy Review}
\newcommand {\nat}      {Nature}
\newcommand {\nphysa}   {Nuclear Physics A}
\newcommand {\nphysb}   {Nuclear Physics B}
\newcommand {\nphysbs}  {Nuclear Physics B Proceedings Supplements}
\newcommand {\physscr}  {Physica Scripta}
\newcommand {\pr}       {Physical Review}
\newcommand {\pra}      {Physical Review A}
\newcommand {\prb}      {Physical Review B}
\newcommand {\prc}      {Physical Review C}
\newcommand {\prd}      {Physical Review D}
\newcommand {\pre}      {Physical Review E}
\newcommand {\PRL}      {Phys. Rev. Lett.}
\newcommand {\prl}      {Phys. Rev. Lett.}
\newcommand {\physrep}  {Physics Reports}
\newcommand {\planss}   {Planetary Space Science}
\newcommand {\pnas}     {Proceedings of the National Academy of Science}
\newcommand {\procspie} {Proceedings of the SPIE}
\newcommand {\pasa}     {Publications of the ASA}
\newcommand {\pasj}     {Publications of the ASJ}
\newcommand {\pasp}     {Publications of the ASP}
\newcommand {\qjras}    {Quarterly Journal of the RAS}
\newcommand {\rmp}      {Reviews of Modern Physics}
\newcommand {\ssr}      {Space Science Reviews}
\newcommand\mathplus{+}

\bibliographystyle{elsarticle-num} 
\bibliography{reference}

\end{document}